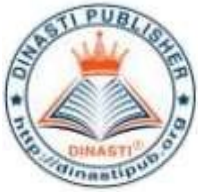 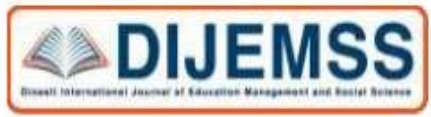

# CREATING EXPERIENCE VALUE TO BUILD STUDENT SATISFACTION IN HIGHER EDUCATION

**Muji Gunarto[1)], Ratih Hurriyati[2)]**
[1)] Faculty of Economics and Business, Universitas Bina Darma, Palembang, Indonesia
[2)] Faculty of Economics and Business, Universitas Pendidikan Indonesia, Bandung, Indonesia

**Abstract:** Higher education products or services received by students are experiential values. The purpose of this study is how to create the values of student experience so that student satisfaction arises. Higher education should now focus on students by creating strong ties with students and alumni. This research was conducted with a survey confirmatory approach. The survey was conducted at 32 universities in South Sumatra Province, Indonesia with a total sample of 357 students. The sampling technique used was stratified random sampling and data analysis using structural equation modeling (SEM) analysis. The results showed that the values of experience in HE were formed through increased co-creation in HE, where students were directly involved in various campus activities. High co-creation shows that there is a stronger attachment of students to HE and higher value of student experience. Co-creation does not directly affect student satisfaction, but it does indirectly affect experience value. If the value of experience is higher, student satisfaction will also be higher.

## INTRODUCTION

Higher Education (HE) in the world is experiencing very rapid development as is the case in Indonesia (Helen & Ho, 2011). Consumer behavior in higher education is typical (Hemsley-Brown, Jane, & Oplatka, 2016), because they make payments in college not directly receiving the benefits, but the value of the experience purchased (Heo & Lee, 2016). Research in the field of higher education marketing is still at an early stage, much research needs to be done both from problem identification and strategic perspectives ((Hemsley-Brown & Oplatka, 2006). Higher education needs to serve customers well so as not to be abandoned (Kotler & Armstrong, 2016). Customer-oriented marketing strategies will be able to provide appropriate value to customers (Gunarto, et al., 2018). Students as college consumers are key actors in contributing to the creation of experiential value (Gunarto, et al., 2018). Higher education provides experience services to students as consumers (Khanna, Jacob, & Yadav, 2014).

The value of higher education services received by students is not short term but is felt in the long term. The product received from a college is the value of experience. Educational services are experiential services in which the active involvement of higher education service providers and consumers or students (Khanna et al., 2014). Students are not directly able to receive the services paid, but the value of experience that will be obtained from starting to become a student to alumni (Gunarto, et al., 2018). The higher level of participation will increase the value of experience (Armstrong & Kotler, 2015; Aygoren & Yilmaz, 2013; Azemi et al., 2016; Mathis, Kim, Uysal, Sirgy, & Prebensen, 2016). The level of student participation in HE is through co-creation. Co-creation will create an interesting and powerful experience for students, so that it causes satisfaction and has an impact on loyalty (Gunarto, et al., 2018; Mathis et al., 2016). Some researchers mentioned that the higher level of participation will increase the value of experience and overall will be able to increase satisfaction (Armstrong & Kotler, 2015; Aygoren & Yilmaz, 2013; Azemi et al., 2016; Mathis et al., 2016). Based on the research gap described above, the researcher is interested in studying how to create experience values that can increase student satisfaction in higher education.

## LITERATURE REVIEW
### *Student Satisfaction*

Satisfaction is defined as the difference between expectations about service and perceptions about the service that it actually receives (Parasuraman, Zeithaml, & Berry, 1988). The level of satisfaction is a function of the difference between perceived performance and expectations, so that customer expectations lie behind why two organizations in the same type of business can be valued differently by their customers. The most important factor for creating customer satisfaction is the performance of the company which is usually interpreted as the quality of the company (Mowen, 2000). Student satisfaction is the feeling of pleasure or disappointment of students that arise after comparing between the services thought about the performance it receives. Customer expectations are formed based on their previous experiences, the advice of friends or colleagues and the promises and information of marketers and competitors (Kotler & Armstrong, 2016; Kotler & Keller, 2016). Student expectations are students' beliefs about what they will receive. Their hopes were shaped by the experience of previous students, comments from friends and acquaintances and promises from the college. These students' expectations over time develop along with the increasing experience of students.

### *Value Experience*

As consumers in higher education, students hope to have a satisfying experience on campus with valuable learning experiences. Management in the classroom becomes valuable for students, so the characteristics of lecturers tend to be the main determinants of student satisfaction in higher education (Gruber et al., 2012). The value of experience in tertiary institutions is created and created together from the start of planning to enter tertiary institutions to the process of studying at the tertiary institution. Researchers recognize the value that customers see as a necessity in consumption and decision making in services (Prebensen, Woo, & Uysal, 2013). Many organizations and universities have gradually begun to apply student-centered approaches such as flexible learning and enhanced technology, so it becomes important to value learning experiences. Learning experience is a process in which students are encouraged to understand actions, reactions, observations, and perceptions of certain situations. This can be achieved by participants by directly sharing whatever their experiences are or by taking part in each activity so as to assimilate the situation in detail and improve their competence (Akhilesh, 2017). The value of experience in learning allows an individual to analyze and observe various approaches applied in diverse situations, both indoors and outdoors (Akhilesh, 2017).

Higher education is a good example of an organization involved in offering an unforgettable student experience. Since students enter as new students, the lecture process to become alumni, places an emphasis on customers and creates more interesting experiences (Gunarto, Wibowo, & Hurriyati, 2016; Gunarto, et al., 2018). Experience value and performance value are fundamentally different (Manschot & Visser, 2011). The value of student experience is the value that is felt, experienced and carried out in relation to higher education, namely the process of teaching, research, community service and others. There are 3 (three) aspects or dimensions that are measured for the value of student experience adopted from Indiana University's CSEQ (The College Student Experiences Questionnaire), namely lecture activities, lecture environment and profits (Patent No. 654321, 1998).

*Co-creation*

One of the broad meanings of co-creation in marketing refers to solving shared problems between buyers and sellers and other actors related to design, production, shipping and purchasing with the aim of creating customer solutions. The starting point for co-creation solutions is the active participation and interaction of buyers with supplier companies in the creation of customer solutions. The key actor who contributes to the creation of value in co-creation is the buyer or customer. This is different from marketing centered where the seller plays a dominant role in value creation and the customer is the recipient of the value created by the seller (Grönroos, 2008; Grönroos & Christian, 1982). Prahalad and Ramaswamy (2004) introduce the concept of co-creation as a unique way to create value for customers. Co-creation of products and services with customers or consumers is a major challenge for marketing managers (Vernette & Hamdi-Kidar, 2013).

Co-creation is a new trend in a business context that aims to bring together all stakeholders, especially customers in different phases of the creation and production of products and services (Ribes-Giner, Perello-Marín, & Díaz, 2016). The process of developing new products always gives birth to new things, where the development of new products is a stage of the process that is full of challenges and high risks (Ranjan & Read, 2016). Educational institutions that implement co-creation can develop competitive strategies that will produce more value for students and will also be difficult to emulate by competitors (Giner & Rillo, 2016).

The most commonly used factors in the co-creation literature are customer participation and involvement, appropriate communication and transparent feedback (Ribes-Giner et al., 2016). Higher education is important to strengthen interaction between students, lecturers, staff and the community with the aim to enhance the learning experience and to achieve student expectations (Pinar, Trapp, Girard, & Boyt, 2011). Students will get more responsibilities and a lecturer becomes a good learning facilitator by implementing co-creation. Some positive results from this interaction are increased program adaptation, learning flexibility and facilitating learning control (Bowden & D'Alessandro, 2011). Another result of co-creation is the positive impact in the curriculum design process, where students have active participation (Bovill, 2014); and can improve knowledge and skills (Ribes-Giner et al., 2016).

## RESEARCH METHODS

The design of this study uses an explanatory survey approach. This research was conducted to determine the clarity of the relationship of a variable and test the hypothesis through data collection in the field. This study was designed to understand or analyze the relationships and effects of exogenous variables on endogenous variables. The unit of analysis in this study is private higher education in South Sumatra Province, while the sample units are final-year students and alumni. The population of this research is all students in HE in South Sumatra Province, amounting to 77,692 people. The sampling technique used in this study was stratified random sampling based on the university form and obtained the number of samples such as Table 1.

**Table 1. Number of Population and Samples**

| No | Form HE | Sum HE | Sum Student | HE Sample | Student Sample |
|---|---|---|---|---|---|
| 1 | University | 14 | 44.342 | 11 | 149 |
| 2 | High School | 53 | 27.356 | 14 | 138 |
| 3 | Academy | 29 | 4.450 | 3 | 24 |
| 4 | Polytechnic | 7 | 1.544 | 4 | 46 |
|   | TOTAL | 103 | 77.692 | 32 | 357 |

The research instrument used a questionnaire with a Likert scale of 7 (seven) categories because of its ease and more popular (Dunn-Rankin, Knezek, Wallace, & Zhang, 2004). Statistical analysis techniques use the Structural Equation Model (SEM) approach, because SEM is a statistical modeling technique that is most common and has been used extensively in behavior science (Gunarto, 2018; Hair, Black, Abin, & Anderson, 2014). Data processing tools that are used with the help of the LISREL program package, because this program is better able to solve complex structural models and have better graphic capabilities than AMOS and EQS (Kline, 1998).

# FINDINGS AND DISCUSSION
## *Profil Responden*
The number of respondents as many as 357 students mostly (44.8%) aged less than 21 years, ages between 21-30 years 44.5%, while ages between 31-40 years there were 6.4% and age over 40 years there were 6, 2%. The sex of the respondent is relatively balanced, there are 54% male and 46% male.

## *Measurement Model*
Analysis of the measurement model is done by confirmatory factor analysis (CFA). Analyzes were performed on each variable to see the ability of the indicator in explaining latent variables. The magnitude of the indicator in explaining latent variables is expressed by the loading factor. Hair et al., (2014) stated that if the factor loading value is greater than 0.5 then the indicator is valid. Reliability tests were carried out with Construct Reliability (CR) and Average Variance Extract (AVE) with the criteria of an instrument or variable declared to have good reliability if CR ≥ 0.7 and AVE ≥ 0.5. If the CR value between 0.6 - 0.7 reliability is still acceptable, provided the indicators have good validity (Hair et al., 2014).

The CFA results for the measurement models for each variable obtained by factor loading values and reliability values as in Table 2.

**Table 2. Value of the Loading Factor and Reliability of Measurement Model**

| Variable | Indicator | Factor loading ($\lambda$) | Square Factor loading ($\lambda^2$) | Error (e) | Construct Reliability (CR) | Average Variance Extract (AVE) |
|---|---|---|---|---|---|---|
| **Co-Creation** | CC1 | 0.81 | 0.66 | 0.34 | 0.95 | 0.65 |
| | CC2 | 0.92 | 0.85 | 0.15 | | |
| | CC3 | 0.74 | 0.55 | 0.45 | | |
| | CC4 | 0.80 | 0.64 | 0.36 | | |
| | CC5 | 0.75 | 0.56 | 0.44 | | |
| | CC6 | 0.80 | 0.64 | 0.36 | | |
| | CC7 | 0.88 | 0.77 | 0.23 | | |
| | CC8 | 0.87 | 0.76 | 0.24 | | |
| | CC9 | 0.70 | 0.49 | 0.51 | | |
| | CC10 | 0.76 | 0.58 | 0.42 | | |
| | CC11 | 0.78 | 0.61 | 0.39 | | |
| **Experience Value** | EV1 | 0.60 | 0.36 | 0.64 | 0.96 | 0.62 |
| | EV2 | 0.67 | 0.45 | 0.55 | | |
| | EV3 | 0.81 | 0.66 | 0.34 | | |
| | EV4 | 0.75 | 0.56 | 0.44 | | |
| | EV5 | 0.72 | 0.52 | 0.48 | | |
| | EV6 | 0.81 | 0.66 | 0.34 | | |
| | EV7 | 0.82 | 0.67 | 0.33 | | |
| | EV8 | 0.71 | 0.50 | 0.50 | | |
| | EV9 | 0.86 | 0.74 | 0.26 | | |
| | EV10 | 0.85 | 0.72 | 0.28 | | |
| | EV11 | 0.81 | 0.66 | 0.34 | | |
| | EV12 | 0.88 | 0.77 | 0.23 | | |

| Variable | Indicator | Factor loading (λ) | Square Factor loading (λ²) | Error (e) | Construct Reliability (CR) | Average Variance Extract (AVE) |
|---|---|---|---|---|---|---|
| | EV13 | 0.85 | 0.72 | 0.28 | | |
| | EV14 | 0.83 | 0.69 | 0.31 | | |
| **Student Satisfaction** | SS1 | 0.80 | 0.64 | 0.36 | **0.95** | **0.74** |
| | SS2 | 0.80 | 0.64 | 0.36 | | |
| | SS3 | 0.86 | 0.74 | 0.26 | | |
| | SS4 | 0.88 | 0.77 | 0.23 | | |
| | SS5 | 0.94 | 0.88 | 0.12 | | |
| | SS6 | 0.89 | 0.79 | 0.21 | | |
| | SS7 | 0.83 | 0.69 | 0.31 | | |

Table 2 shows all the indicators in the measurement model of each variable are valid, because it has a loading factor value (λ) of more than 0.5. Judging from the value of CR and AVE in each latent variable also shows reliable, because CR is greater than 0.7 and AVE is more than 0.5. This shows that the indicators in the measurement model for co-creation, experience value and student satisfaction variables are valid and reliable. All indicators are able to explain the construct variables well.

*Full Model Structure*

After confirmatory factor analysis (CFA) of each variable, then a full model analysis is performed to form a fully fit structural model. Estimation results for the full structural model analysis are shown as in Figure 1.

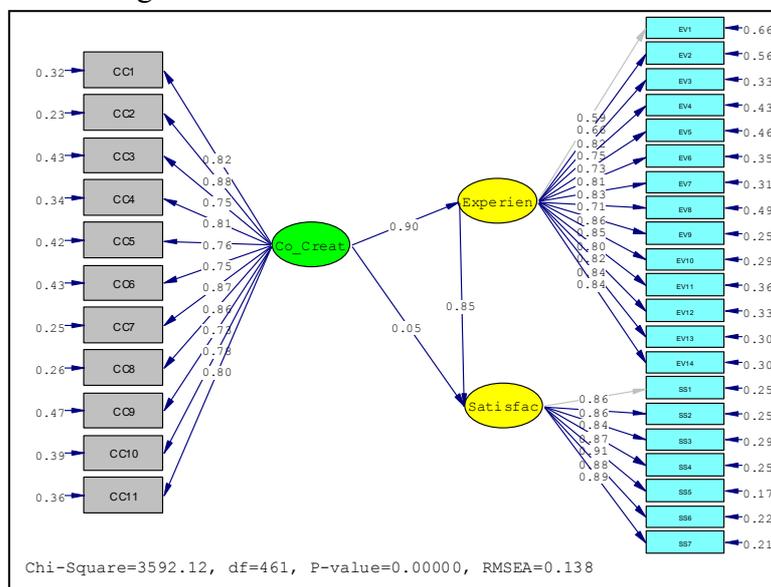

**Figure 1. Full Structural Model Estimation Results**

Figure 1 shows the magnitude of the parameter values in the relationships between the latent variables and the magnitude of the loading factor values for each indicator. Based on the parameter values, there is a positive relationship between co-creation with experience value and student satisfaction. Student satisfaction is strongly influenced by the value of experience (0.85), while co-creation only affects 0.05. The value of co-creation also has a very high positive influence (0.90) on the value of experience for students.

The test results for all parameters are indicated by the t-value which looks like in Figure 2.

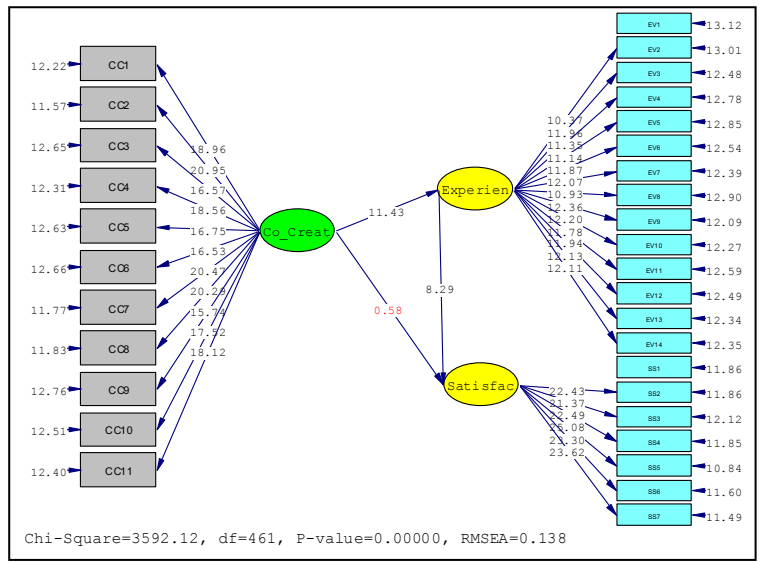

**Figure 2. Full Structural Model Test Results**

Figure 2 shows the results of testing all parameters, namely for the measurement model and structural model. Based on Figure 2 shows that all indicators in the measurement model are significant, because the t-value obtained is greater than 1.96, while the structural test results for the relationship between latent variables are not significant. The results of the structural model fill test are shown in Table 3.

**Table 3. Testing Results of Relationships Between Latent Variables**

| No. | Endogen Variable | | Exogen Variable | Estimate | S.E. | t- Value | $R^2$ |
|---|---|---|---|---|---|---|---|
| Model (1) | Experience value | <--- | Co-creation | 0,90 | 0,08 | 11,43[*)] | 0,82 |
| Model (2) | Student satisfaction | <--- | Experience value | 0,85 | 0,10 | 8,29[*)] | 0,80 |
|  | Student satisfaction | <--- | Co-creation | 0,05 | 0,08 | 0,58[ns)] |  |

Note: [*)] = Significance, [ns)] = Not Significance

*Discussion*

The mathematical equation in the sub-structure model (1) is obtained as follows.

　　　　***Experience = 0.90*Co_Creation　　(R² = 0.82)***　　　　　　　　　…(1)

Model (1) shows that co-creation has a positive effect on experience value of 0.90 and is statistically significant at 5% significance level. If the co-creation conducted by universities is getting better, the value of the experience felt by students is also higher. The model is able to explain variations in existing data by 82%.

The mathematical equation of the sub-structure model (2) is obtained as follows.

　　　***Satisfaction = 0.85*Experience + 0.045*Co_Creation, (R² = 0.80)***　　　…(2)

Model (2) shows that co-creation has a positive effect on student satisfaction by 0.05, but statistically it is not significant at 5% significance level (table 3). The experience value had a positive effect on student satisfaction by 0.85 and statistically significant at the 5% level. This means that the better the value of student experience, the higher student satisfaction. Model (2) is able to explain the variation of sample data by 80%.

The results of the two models above indicate that co-creation does not directly affect student satisfaction, but has an indirect effect, namely through the value of experience. The existence of a very strong influence of co-creation on the value of experience that has an impact on student satisfaction shows that the value of experience is a good intervening variable for the relationship between co-creation with student satisfaction. The absence of co-creation relations with student satisfaction illustrates that many aspects of co-creation still need to be improved, because the value of co-creation in higher education includes more aspects and actors (Díaz-Méndez & Gummesson, 2012).

The value of student experience in tertiary institutions can be built through engagement between students and tertiary institutions, namely through co-creation. The co-creation built by HE makes students the main actors in contributing to the creation of value, so that in the co-creation at HE students play a very important role. The value of co-creation in higher education is an interactive dialogue that involves the ability and willingness to act on both sides (Prahalad and Ramaswamy, 2004). Bowden & D'Alessandro (2011) explained that HE with a better level of co-creation would increase the value of the experience for students compared to using a traditional approach. The results of other studies state that co-creation that is run effectively can increase the value of experience for students (Akhilesh, 2017; Dean, Griffin, & Kulczynski, 2016; Elsharnouby, 2015; Fuller, Hutter, & Faullant, 2011; Mathis et al., 2016; Prahalad & Ramaswamy, 2004).

**CONCLUSION AND SUGESTION**

Student experience scores are a measure of service received by students at HE. The higher the value of experience received, the lower the fees paid by students. The values of experience in HE can be formed through increased co-creation in tertiary institutions, where students are directly involved in various activities on campus both in the classroom and outside the classroom. High co-creation shows that there is a stronger attachment of students to HE and higher value of student experience. Co-creation does not directly affect student satisfaction, but does not directly affect the value of experience. If the value of experience is higher, the higher student satisfaction.

Higher education should not only focus on academic and tertiary reputation, but rather focus on the values of student experience. Involve various university activities such as curriculum development with students and alumni. Create the value of student experience by joint creation between lecturers and students. Invite students to dialogue for activities to be carried out by the college.